\documentclass[preprint,showpacs,preprintnumbers,showkeys,aps,prd,a4paper,byrevtex]{revtex4}

\usepackage[utf8]{inputenc}

\usepackage[T1]{fontenc}        % Tries to use Postscript Type 1 Fonts for better rendering
\usepackage[intlimits]{amsmath} % Provides all mathematical commands
\usepackage{amssymb,amscd}

\usepackage{hyperref}           % Provides clickable links in the PDF-document for \ref
\usepackage{graphicx}
\graphicspath{{./figs/}}

\usepackage[ugly]{units}        % Allows you to type units with correct spacing and font style. Usage: $\unit[100]{m}$ or $\unitfrac[100]{m}{s}$

\begin{document}

%%%%%%%%%%%%%%%%%%%%%%%%%%%%%%%%%%%%%%%%%%%%%%%%%%%%%%%%%%%%%%%%%%%%%%%%%%%%%%%%%
%opening
\title{Production of exotic particles in electron-positron collisions}
\author{W.~K. Sauter}
\email{werner.sauter@ufpel.edu.br}
\affiliation{Grupo de Altas e Médias Energias \\ Departamento de Física - Instituto de Física e Matemática \\ Universidade Federal de Pelotas}

\date{\today}

\begin{abstract}
 In this work is presented several results for cross sections of exotic particles production in electron-positron collisions in future linear electron-positron colliders. It is shown that the central production of exotic leptons, dilatons, radions, axions and technipions in this type of collision can be measurable for the predicted kinematic parameters for linear colliders.
\end{abstract}

\preprint{Version 1.0}
\maketitle

\section{Introduction}
\begin{figure}[ht] \label{fig:proc}
\begin{center}
\includegraphics[width=0.4\textwidth]{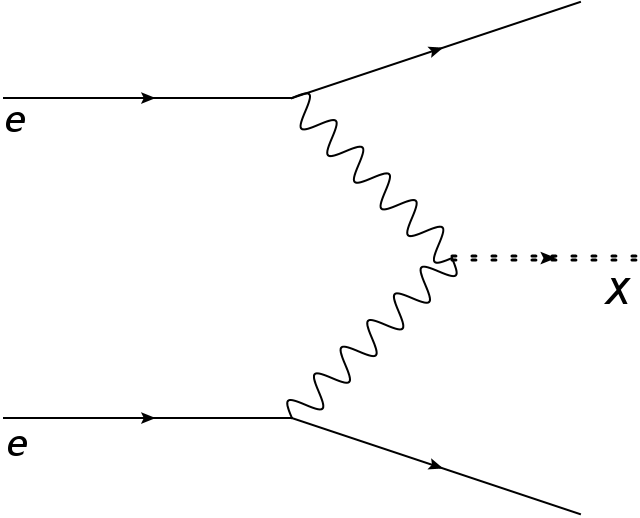}\includegraphics[width=0.35\textwidth]{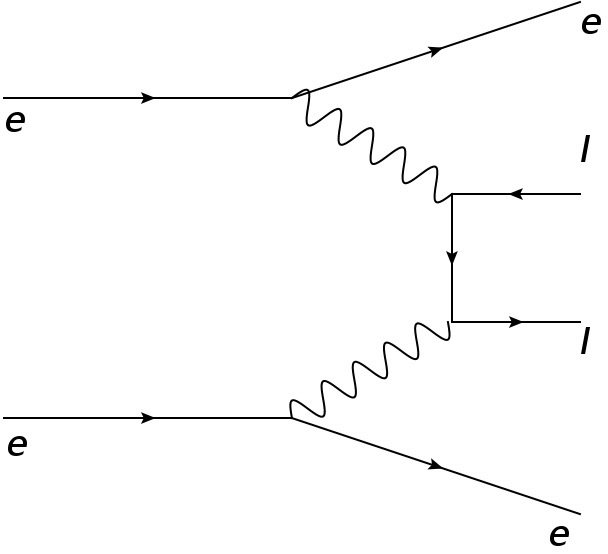}
\end{center}
 \caption{Feynman diagram for the process of central production of exotic particle in $ee$ collisions (left) and dileptons (right).}
\end{figure}

With the advent of the next generation linear colliders, opens the possibility of studying the production of particles not predicted in the Standard Model in a new kinematic regime. One of the advantages of using electrons and positrons as projectiles is its absence of a known substructure, in contrast to hadrons. Due to this simplification, it is possible to study the direct production of states beyond the Standard Model (BSM). The ideia is use the formalism of peripheral collisions to calculate total cross sections in electron positron collisions, $e^- e^+ \rightarrow e^- \mathcal{X} e^+$, where $\mathcal{X}$ is a beyond Standard Model particle (see fig. (\ref{fig:proc}). The following different possibilities for this exotic states are: radion, dilaton, technipion and axion. The BSM physics opportunities in future linear electron colliders are described in \cite{1812.02093v2} for CLIC and in \cite{1511.00405v1} for ILC.

The central production of resonances is a very good process to search and study the exotic particles states due his advantages. In previous works, we already obtain results of magnetic monopoles\cite{Reis:2017rvb} and radion\cite{goncalves2010RadproexcproCERLHC} and dilaton\cite{Goncalves:2015oua} production in hadron-hadron collisions. An advantage of this process is the measurement of the projectiles after the interaction, enabling the measurement of the energy deposited in the detector of the produced central state or by its decay products. Another advantage is its very clean experimental signature on the detector: two leptons separated by a rapidity interval (angle) of a central activity on the detector.

\section{Theory}
We use the simple model in equivalent photon approximation~\cite{nystrand2005Eleintnucprocol} to calculate the total cross sections, valid in diferent cases: electron-electron, proton-proton and ion-ion collisions. The expression for cross-section reads
\begin{equation} \label{eq:photequiv}
\sigma_\mathrm{tot} = \int_{M_{\mathcal{X}}^2/S}^1\!dx_1\ f_A(x_1) \int_{M_{\mathcal{X}}^2/x_1S}^1\!dx_2\ f_B(x_2) \sigma_{\gamma\gamma\mathcal{X}}(\hat{s})
 \end{equation}
where $M_{\mathcal{X}}$ is the mass of the central produced system, $S$ is the center-of-mass energy of the projectiles, $x_i$ is the fraction of the energy of the photon $i$, $\hat{s} = x_1 x_2 s_{NN}$, $f(x)$ is the photon energy spectrum (or photon flux) produced by a charged particle and $\sigma_{\gamma\gamma\mathcal{X}}(\hat{s})$ is the cross section of production of a $\mathcal{X}$ state by a fusion of a pair of photons.

For the electron collisions, we use two different expressions for the photon energy spectrum: the first one from Frixione\cite{frixione1993ImpWeiappele-procol},
\begin{equation} \label{eq:frix}
 f_{e}^{F}(x) = \frac{\alpha_{elm}}{2\pi} \bigg\{ 2m_{e}^{2}x \left[ \frac{1}{q^2_{max}} - \frac{1}{q^2_{min}}\right] + \frac{1+(1-x)^2}{x}\log\left(\frac{q^2_{min}}{q^2_{max}}\right) \bigg\}
\end{equation}
where 
\[ 
 q^2_{max} = -\frac{m_{e}^{2}x^2}{1-x},\quad q^2_{min} = -\frac{m_{e}^{2}x^2}{1-x} - E^2(1-x)\theta_c^2
\]
with $m_e$ being the electron mass, $E$ the energy beam, and $\theta_{c} = \unit[30]{mrad}$; and the second one photon flux comes from Budnev \textit{et al.}\cite{budnev1975TwophoparpromecPhyproAppEquphoapp},
\begin{multline} \label{eq:bud}
 f_{e}^{B}(x) = \frac{\alpha_{elm}}{\pi}\frac{1}{xE^2} \left[ \left(E^2 + E^{\prime 2} \right) \ln \frac{2e}{m_e} - EE^\prime + \frac{1}{2}(E-E^\prime)^2 \ln\frac{E^\prime}{E - E^\prime} \right. \\
 \left. + \frac{1}{2}(E+E^\prime)^2 \ln\frac{E^\prime}{E + E^\prime} \right],
\end{multline}
where $E^\prime = E(1-x)$ is the energy of the scattered photon.

\begin{figure}[ht] \label{fig:lum}
\begin{center}
\includegraphics[width=0.6\textwidth]{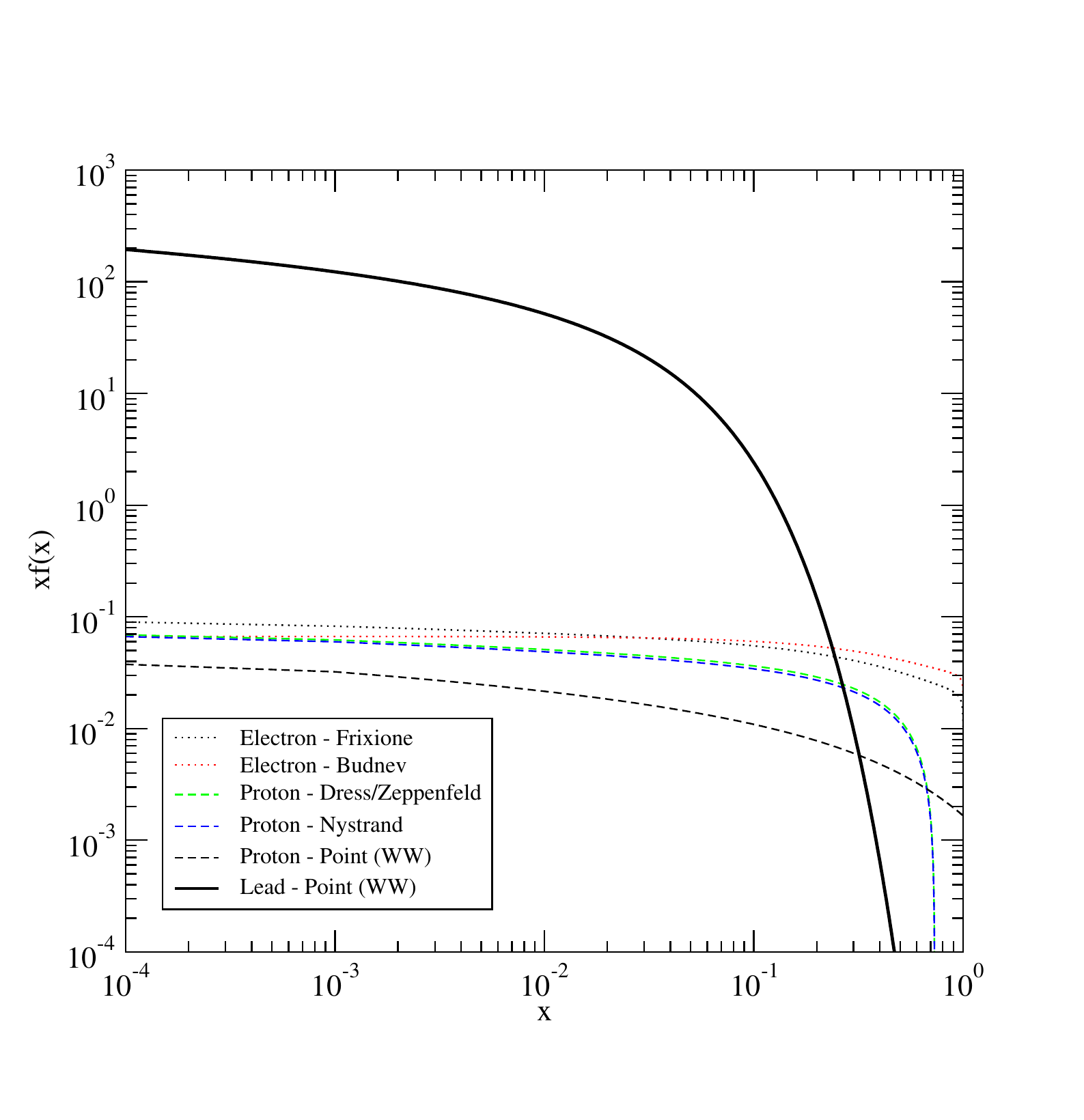}
\end{center}
 \caption{Photon luminosities from different projectiles (electron, proton and ions) as a function of the energy fraction. See the text for explanation of the different cases. }
\end{figure}

A comparison with the luminosities of the electron, proton and nuclei is displayed in the figure (\ref{fig:lum}). In this figure, we plot the electron photon flux from the above expressions, eq. (\ref{eq:frix}) and eq. (\ref{eq:bud}); in the proton case, we plot the expressions from Dress and Zeppenfeld\cite{drees1994gamgamprohigppcol}, from Nystrand\cite{nystrand2005Eleintnucprocol} and the result of Wëiszacker and Williams for a point charge photon flux (see the expression in 
\cite{nystrand2005Eleintnucprocol}). Note the quite different behavior depending of the projectile: ions (lead, in the case) produce numerous photons at small $x$ values, whereas, in large $x$ values, the larger photon flux comes from electron projectiles, favoring the production of states with great masses. Regarding electron luminosities, both expressions result in very close photon luminosities.

The last term of the total cross section of production of central state is the cross section of the process $\gamma\gamma \rightarrow \mathcal{X}$, given by
\begin{equation}
 \sigma_{\gamma\gamma\mathcal{X}} = 8\pi(2J+1)\frac{\Gamma_{\mathcal{X}\gamma\gamma} \Gamma_{tot}}{(W_{\gamma\gamma}^2 - M_\mathcal{X}^2)^2 + M_\mathcal{X}^2\Gamma_{tot}^2}
\end{equation}
where $J$ is the spin of the produced state, $\Gamma_{\mathcal{X}\gamma\gamma}$ is the partial decay width of $\mathcal{X}$ in two photons, $\Gamma_{tot}$ is the total decay width, and $W_{\gamma\gamma}$ is the two photons CM energy. In the case of a narrow resonance ($\Gamma_{tot} \ll M_\mathcal{X}$) the above result, using a Dirac delta function, reads 
\begin{equation}
 \sigma_{\gamma\gamma\mathcal{X}} = 4\pi^2 (2J+1) \frac{\Gamma_{\mathcal{X}\gamma\gamma}}{M^2_\mathcal{X}}\delta(W_{\gamma\gamma}-M_\mathcal{X}).
\end{equation}
Is easy to obtain the following expression (using proprieties of Dirac delta function) from Eq. (\ref{eq:photequiv}),
\begin{equation} \label{eq:final}
 \sigma_\mathrm{tot}^{\gamma\gamma\mathcal{X}} = \frac{8\pi^2 (2J+1)}{M^2_\mathcal{X}S} \int_{M_{\gamma\gamma}^2/S}^1\!\frac{dx_1}{x_1}\ f_e(x_1) \Gamma_{\mathcal{X}\gamma\gamma} f_e\left(\frac{M_R^2}{x_1 S}\right)
\end{equation}
This is the final expression for the calculation of the cross-section for $ee \rightarrow e\mathcal{X}e$. In the following, we consider some particular cases of central produced particles to obtain the decay width in two photons, the remaining part of the cross section, Eq. (\ref{eq:final}).

\section{Results}

In this section, we will calculate the particle production considering photon - photon interactions for electron-positron collisions at CLIC/ILC energies. In particular, we will extend the results for particle production in hadron collisions. All the cross sections are obtained using the planned energies for electron-positron colliders: $\sqrt{s} = \unit[0.5]{TeV}$ for ILC; $\sqrt{s} = \unit[1.5]{TeV}$ and $\sqrt{s} = \unit[3.0]{TeV}$ for CLIC. We also use the two different photon luminosities. In next subsections, we consider the particular cases of central produced systems.

%%%%%%%%%%%%%%%%%%%%%%%%%%%%%%%%%%%%%%%%%%%%%%%%%%%%%%%%%%%%%%%%%%%%%%%%%%%%%%%%%%
\subsection{Dilepton production}
\begin{figure}[ht] \label{fig:dilep}
\begin{center}
\includegraphics[width=0.6\textwidth]{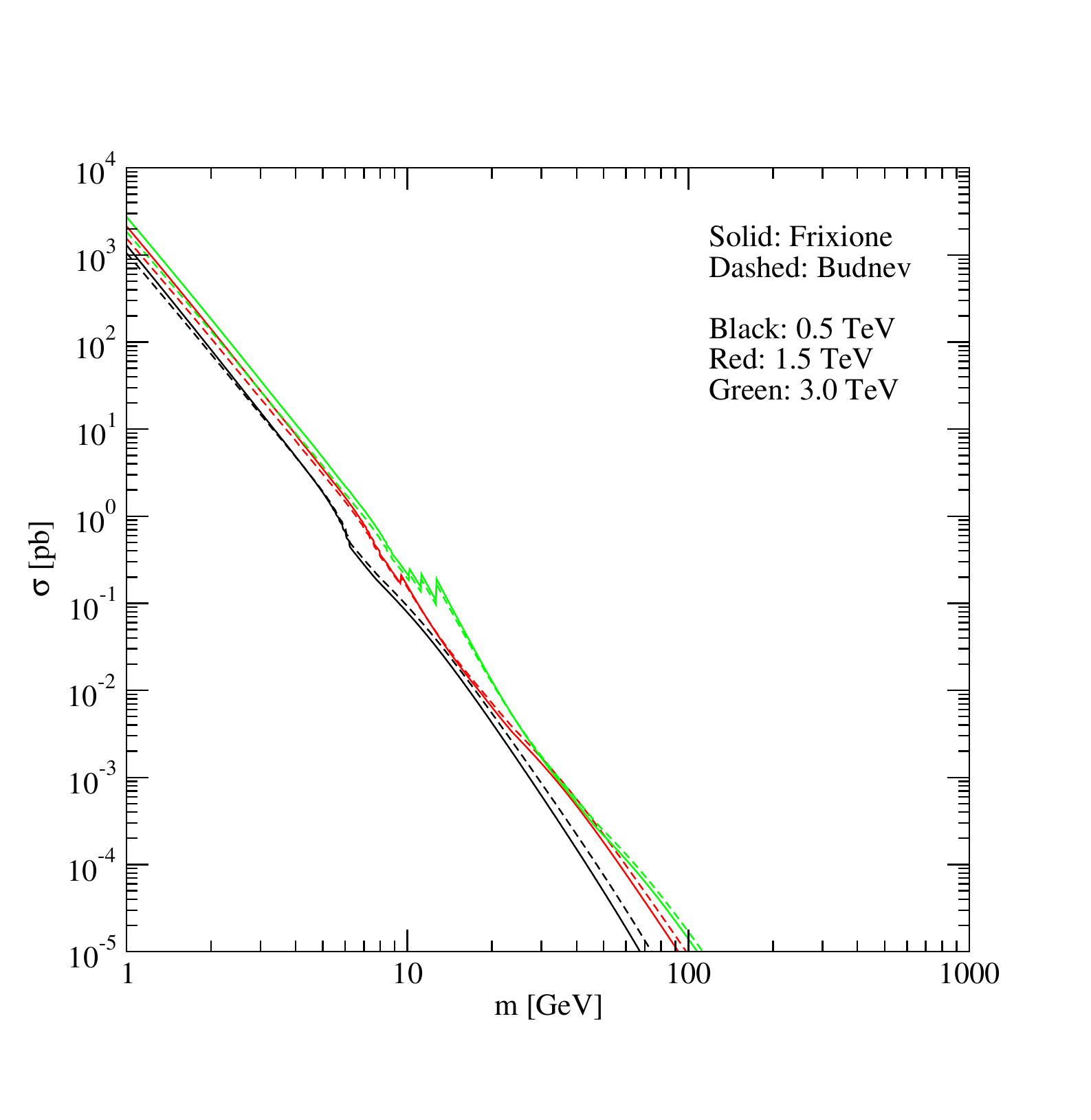}
\end{center}
 \caption{Cross section for the production of a pair of leptons: \protect{$e^+e^-\rightarrow e^+l^+l^-e^-$} for different energies and photon luminosities, see legend in figure.}
\end{figure}

Before considering more complex exotic states, let's consider the production of dileptons. The dilepton production is one of the ``standard candle'' process of central production. This process ($\gamma\gamma \rightarrow \mu^+\mu^-$) have a well known cross section, given by the Breit-Wheeler formula (see \cite{greinerqed} and \cite{KlusekGawenda:2010kx}),
\begin{multline}
\sigma_{\gamma\gamma\rightarrow\mu\mu}(W_{\gamma\gamma}) = \frac{4\pi\alpha^2_\mathrm{em}}{W_{\gamma\gamma}^2} \left\{ 2 \ln \left[ \frac{W_{\gamma\gamma}}{2m_\mu} \left(1 + \sqrt{1 - \frac{4m_\mu^2}{W_{\gamma\gamma}^2}} \right)  \right] \left(1 + \frac{4m_\mu^2}{W_{\gamma\gamma}^2} - \frac{8m_\mu^2}{W_{\gamma\gamma}^4} \right) \right. \\
 \left. - \sqrt{1-\frac{4m_\mu^2}{W_{\gamma\gamma}^2}} \left( 1 + \frac{4m_\mu^2}{W_{\gamma\gamma}^2} \right) \right\}.
\end{multline}
The (preliminary) results for the energies planned in ILC/CLIC is shown in Fig. (\ref{fig:dilep}). In the result obtained, we consider a mass range with a high upper limit to consider, in addition to light particles, other particles such as charginos\cite{godunov2020quasistable}. 

\subsection{Radion}
Several open problems in Physics, such as the hierarchy problem, have a possible solution in the existence of extra dimensions. There are several extra dimension scenarios in the literature. In recent years, the scenario proposed by Randall and Sundrum (RS)\cite{randall1999Larmashiesmaextdim}, in which there are two (3+1)-dimensional branes separated in a 5th dimension, has attracted a great deal of attention. This model predicts a Kaluza-Klein tower of gravitons and a graviscalar, called radion, which stabilize the size of the extra dimension without fine tuning of parameters and is the lowest gravitational excitation in this scenario. The mass of radion is expected to be $\approx$ TeV which implies that the detection of the radion will be the first signature of the RS model.

In this paper we extend the previous studies for exclusive processes\cite{goncalves2010RadproexcproCERLHC} for electron colliders. The signal would be a clear one with a radion tagged in the central region of the detector accompanied by two electrons. In contrast to the inclusive production, which is characterized by large QCD activity and backgrounds which complicate the identification of a new physics signal, the exclusive production will be characterized by a clean topology mediated by colorless exchanges. 

 The $\gamma \gamma \rightarrow \Phi$ cross section can be expressed as follows
\begin{equation}
 \sigma_{\gamma \gamma \rightarrow \Phi} = \frac{8 \pi^2}{m_{\phi}^3} \Gamma(\Phi \rightarrow \gamma\gamma) \,\,.
\end{equation}
The partial decay width of radion into two photons was calculated in \cite{bae2000PheradRanscecol,cheung2001PheradRansce} and is given by:
\begin{multline}
 \Gamma(\Phi \rightarrow \gamma\gamma)   =   \frac{\alpha_{em}^2 m_{\Phi}^3}{256 \pi^3 \Lambda_{\Phi}^2} \left\lbrace  - \frac{22}{6} - [2 + 3x_W + 3x_W(2-x_W)f(x_W)] + \right. \\
 \left. \frac{8}{3} x_t [1+ (1-x_t)f(x_t)]  \right\rbrace ^2,
\end{multline}
with $m_\Phi$ is the radion mass, $\Lambda_{\Phi} = \langle \Phi \rangle \approx {\cal{O}}(v)$ ($v$ the VEV of the Higgs field) determines the strength of the coupling of the radion to the standard model particles, $x_i = 4m_i^2/m_\Phi^2$ (with $i = W,t$ denoting the $W$ boson and the top quark) and the auxiliary function $f(z)$ being given by
\begin{equation} \label{eq:faux}
f(z) = \begin{cases}
\left[ \sin^{-1} \left(\frac{1}{\sqrt{z}} \right ) \right ]^2, & z \ge 1 \\
-\frac{1}{4} \left[ \log \frac{1+\sqrt{1-z}}{1-\sqrt{1-z}} - i \pi \right ]^2. & z <1
       \end{cases}
\end{equation}

\begin{figure}[ht] \label{fig:rad}
\begin{center}
\includegraphics[width=0.6\textwidth]{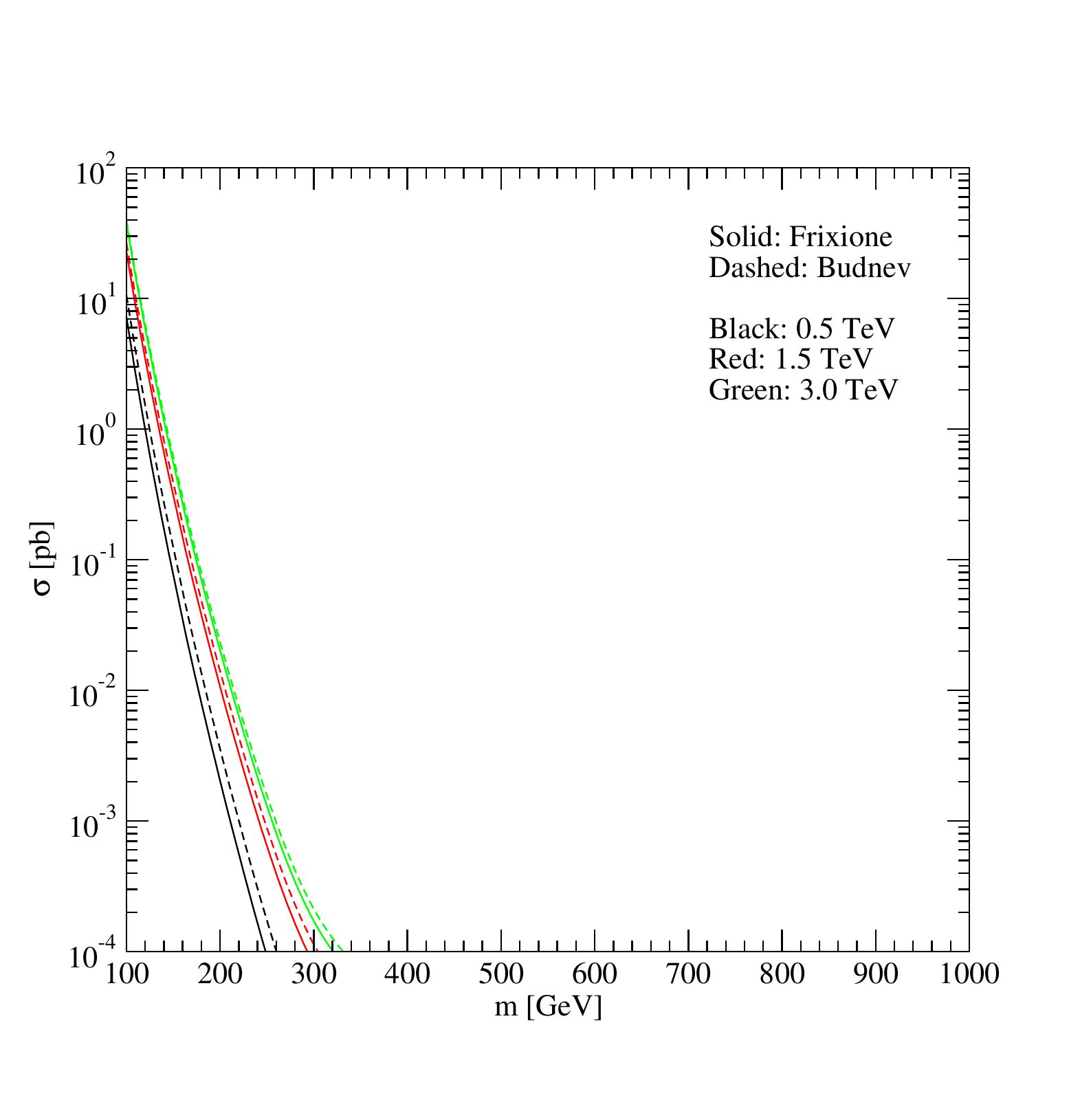}
\end{center}
 \caption{Cross section for the production of radion for different energies and photon luminosities, see legend in figure.} 
\end{figure}

The result for cross section of production of radion as function of mass is displayed in the Fig. (\ref{fig:rad}).

\subsection{Dilaton}
The prediction of the existence of new scalar particles is a characteristic of several candidate theories beyond the standard model (SM) (See e.g. Ref. \cite{ScienceDirect:S037015731200083X}). One of these particles is the dilaton, denoted as $\chi$, which is predicted to appears as a pseudo-Nambu-Goldstone boson in spontaneous breaking of scale symmetry \cite{clark1987ProDil}. In the particular scenario in which electroweak symmetry is broken via strongly coupled conformal dynamics, a neutral dilaton is expected with a mass below the conformal symmetry breaking scale $f$ and couplings to standard model particles similar to those of the SM Higgs boson. The searching of the dilaton in {\it inclusive} proton - proton collisions at LHC energies motivated a lot of work, with particular emphasis in the discrimination of the dilaton from the SM Higgs signals\cite{barger2012DilLHC, barger2012DifHigbosdilradhadcol, coleppa2012DilconLHCpro, coriano2013DilintanobrescainvStaMod, chacko2013Res125GeVHigorDil, bellazzini2013HigDil, jung2014HigsysconLHCHigdat}. In this paper we extend the previous study for {\it exclusive} processes\cite{Goncalves:2015oua}.

The partial decay width of the dilaton into two photons,  $\Gamma_{{\chi} \rightarrow \gamma \gamma}$, was calculated in \cite{coleppa2012DilconLHCpro} and is given by:
\begin{equation} \label{eq:lamgamgam}
\Gamma_{{\chi}\rightarrow \gamma\gamma}(M_{\chi}) = {\cal{C}}_{ \gamma } \frac{\upsilon^2}{f^2} \frac{G_F \alpha_\mathrm{em} M_{\chi}^3}{128 \sqrt{2} \pi^3} \lvert f(\tau_W) + \sum_f N_c Q_f^2 f(\tau_f) \rvert^2
\end{equation}
where $\upsilon = \unit[246]{GeV}$ is the scale of electroweak symmetry breaking, $f$ is the energy scale of conformal scale, $\alpha_\mathrm{em}$ is the electromagnetic coupling constant, $N_c$ is the number of colors, $Q_f$ is the fermion charge. Moreover, the coefficient ${\cal{C}}_{ \gamma }$ is given by \cite{coleppa2012DilconLHCpro}
\begin{equation}
{\cal{C}}_\gamma = \frac{\lvert -b_{EM} + \sum_{i=f,b}N_{c,i} Q_i^2 F_i(\tau_i) \rvert^2}{\lvert \sum_{i=f,b}N_{c,i} Q_i^2 F_i(\tau_i) \rvert^2}
\end{equation}
where $b_{EM} = - 11/3$, the sum runs over fermions ($f$) and bosons ($b$), $N_{c,i}$ is the color multiplicity number ($N_{c,i}=1$ for bosons and leptons and $N_{c,i}=3$ for quarks) and $Q$ is the electric charge in units of $e$.

\begin{figure}[ht] \label{fig:dil}
\begin{center}
\includegraphics[width=0.6\textwidth]{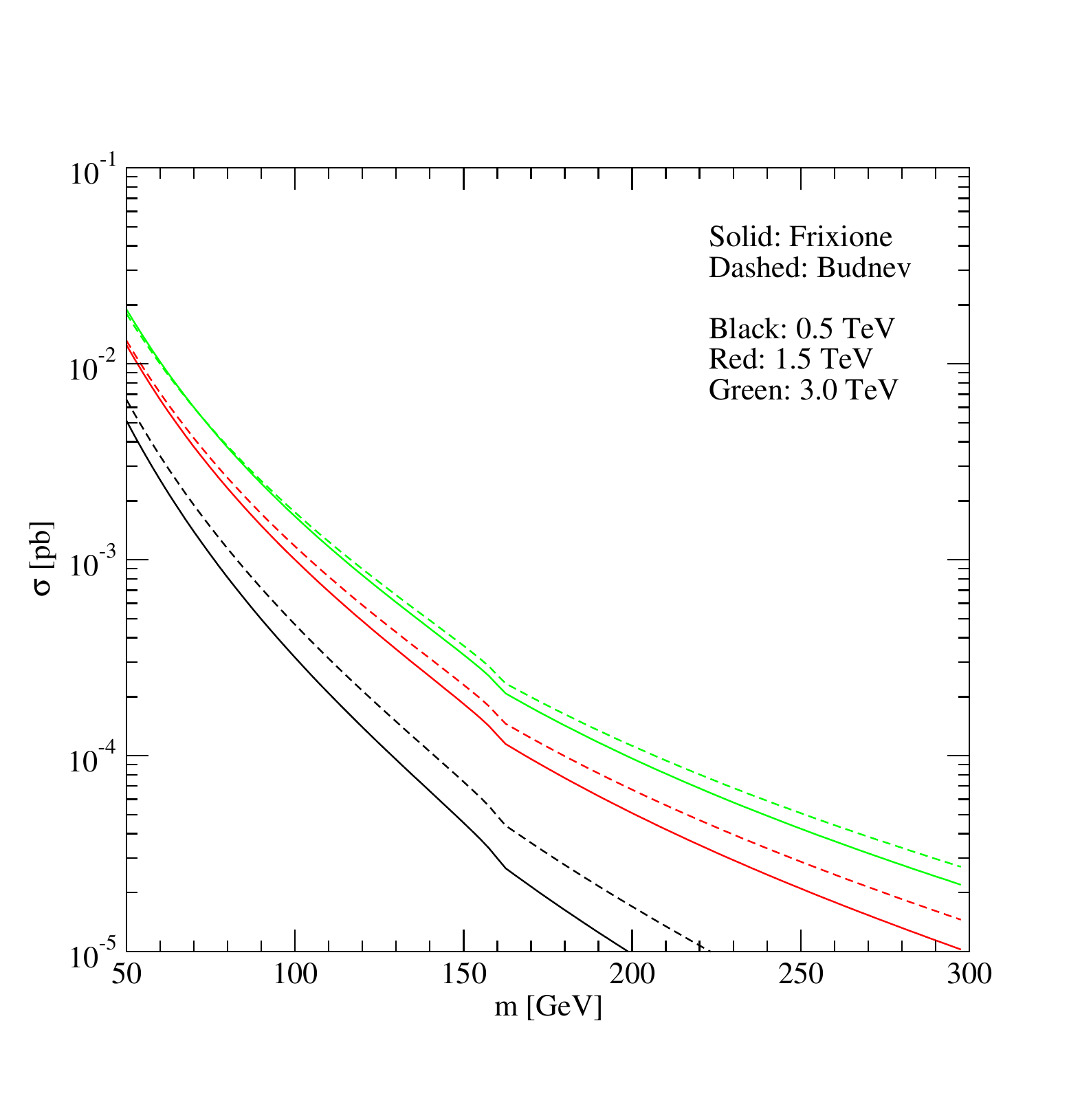}
\end{center}
 \caption{Cross section for the production of dilaton, for different energies and photon luminosities, see legend in figure.} 
\end{figure}

The result for cross section of production of dilaton as function of mass is displayed in the Fig. (\ref{fig:dil}).

\subsection{Axion}
The QCD axion was proposed to solve the strong-CP problem in QCD and is a candidate to constitute the cosmological dark matter. The axion have a rich phenomenology and for an introduction for it see \cite{Ebadi:2019gij, Bauer:2018uxu, Bauer:2017ris, Rybka:2017swh}. In few words, the strong-CP problem is the extremely small neutron electric dipole model, that indicates the fine-tuned cancellation of CP violation in QCD. Among other solutions, the Peccei-Quinn one dynamically cancels CP-violation in QCD but introduces the axion, a pseudoscalar particle.

The central production of axions or ALPs (axion likes particles) in hadronic collisions by photon or gluon fusion in central exclusive processes is analyzed in [???]. Recently, \cite{2003.01978v1} analyzed the virtual production of axions in CLIC. 

\begin{figure}[ht] \label{fig:axi}
\begin{center}
\includegraphics[width=0.6\textwidth]{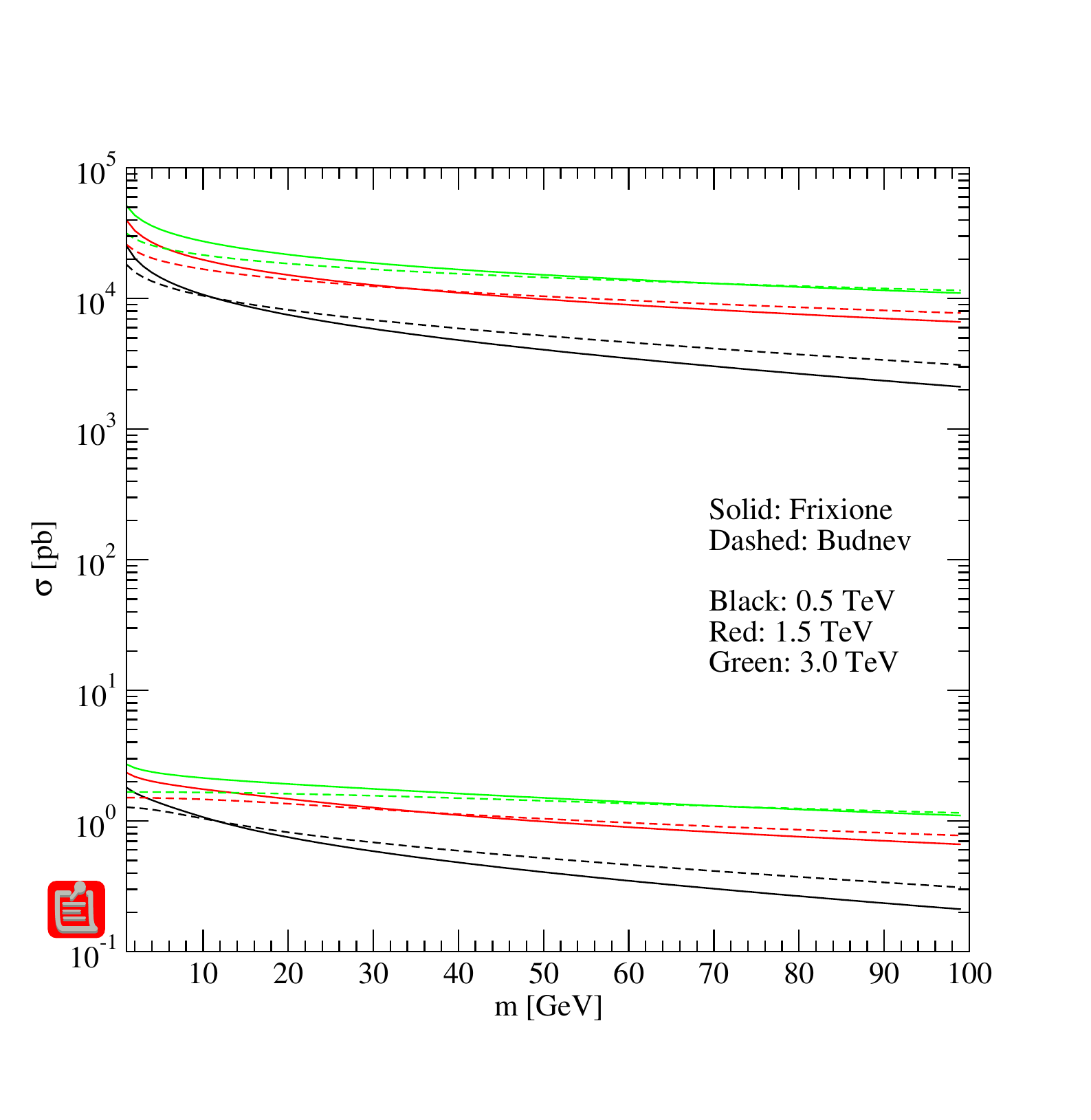}
\end{center}
 \caption{Cross section for the production of axion, for different energies and photon luminosities, see legend in figure. In upper curves, $1/f^2 = 10^{-2}$ and in lower curves, $1/f^2=10^{-4}$.} 
\end{figure}

Here the results are obtained using the following expression for decay width \cite{Ebadi:2019gij, Knapen:2016moh, Baldenegro:2018hng},
\begin{equation} \label{eq:lgax}
\Gamma(a\rightarrow\gamma\gamma) = \frac{1}{4\pi}\frac{1}{f^2} m_a^3, 
\end{equation}
where $\unit[1]{GeV}<m_a<\unit[100]{GeV}$ and $1/f^2 = \unit[10^{-2}]{GeV^{-1}}, \unit[10^{-4}]{GeV^{-1}}$ are related with the axion coupling with photons.
The result for cross section of production of dilaton as function of mass is displayed in the Fig. (\ref{fig:axi}).

\subsection{Technipions}
 Light exotic states are predicted by high-scale strongly coupled dynamics, known as technicolor. In \cite{lebiedowicz2014SeatecexcprodiplarinvmasLHC} the theory and results for the central exclusive production of technicolor particles, in particular, a neutral technipion, in hadron collisions in LHC are presented. One of the results of \cite{lebiedowicz2014SeatecexcprodiplarinvmasLHC} is the dominance of the diphoton decay of technipion for light masses. We use this fact to obtain the cross section in electron-electron collisions. Using the previous formalism, the two-photon technipion decay width reads
\begin{equation} \label{eq:tcp} 
 \Gamma(\tilde{\pi}^0\rightarrow\gamma\gamma) = \frac{m_{\tilde{\pi}}^3}{64\pi} \left| \frac{4\alpha_{elm}g_{TC}}{\pi} \frac{M_{\tilde{Q}}}{m_{\tilde{\pi}}^2} \arcsin^2\left( \frac{m_{\tilde{\pi}}} {2M_{\tilde{Q}}} \right) \right|^2  
\end{equation}
where $m_{\tilde{\pi}}$ is the technipion mass, $\alpha_{elm}$ is the fine structure constant, $g_{TC} = 10$ is a effective coupling and $M_{\tilde{Q}}$ is the mass of techniquark.

\begin{figure}[ht] \label{fig:tech}
\begin{center}
\includegraphics[width=0.45\textwidth]{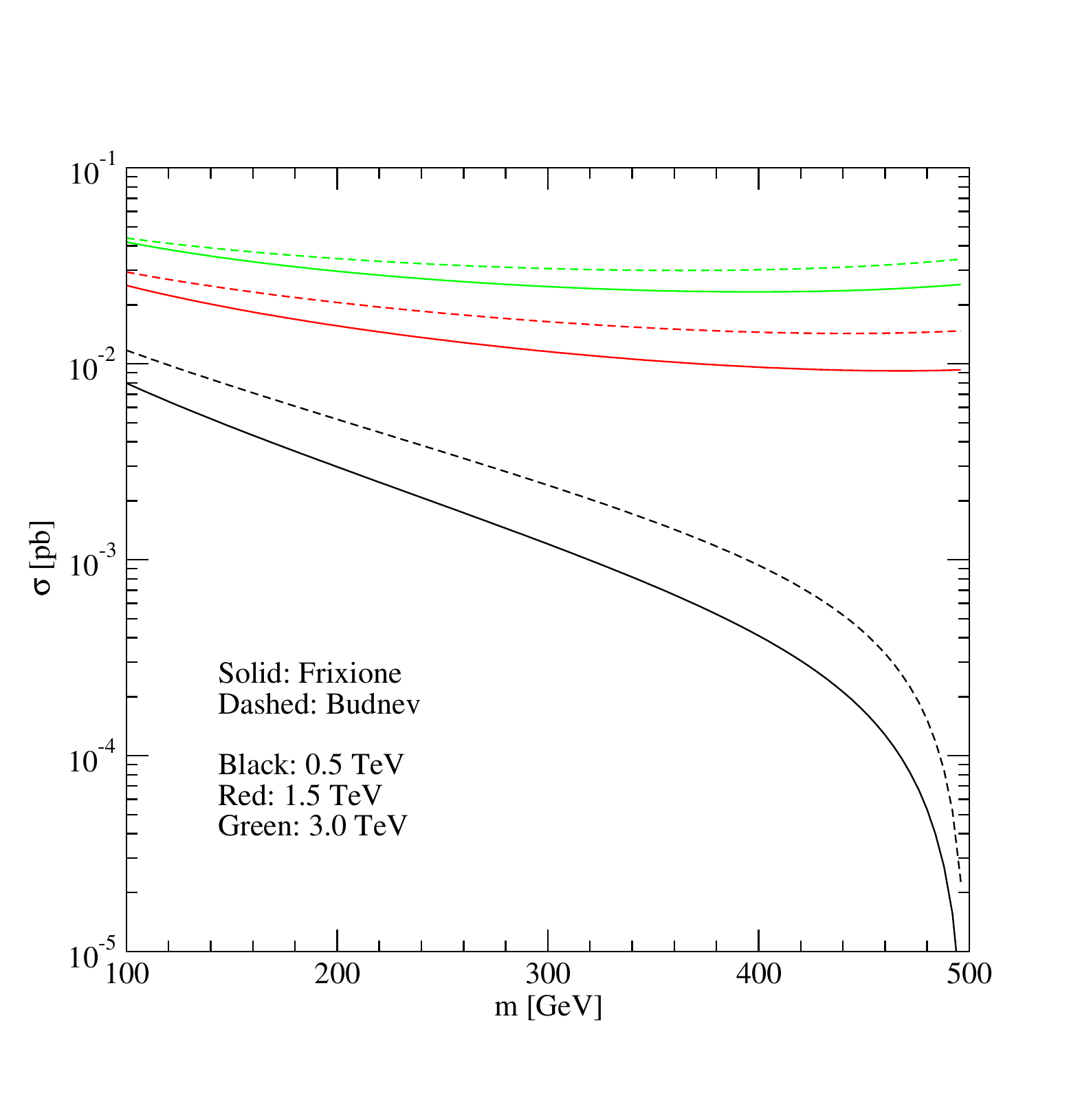}\includegraphics[width=0.45\textwidth]{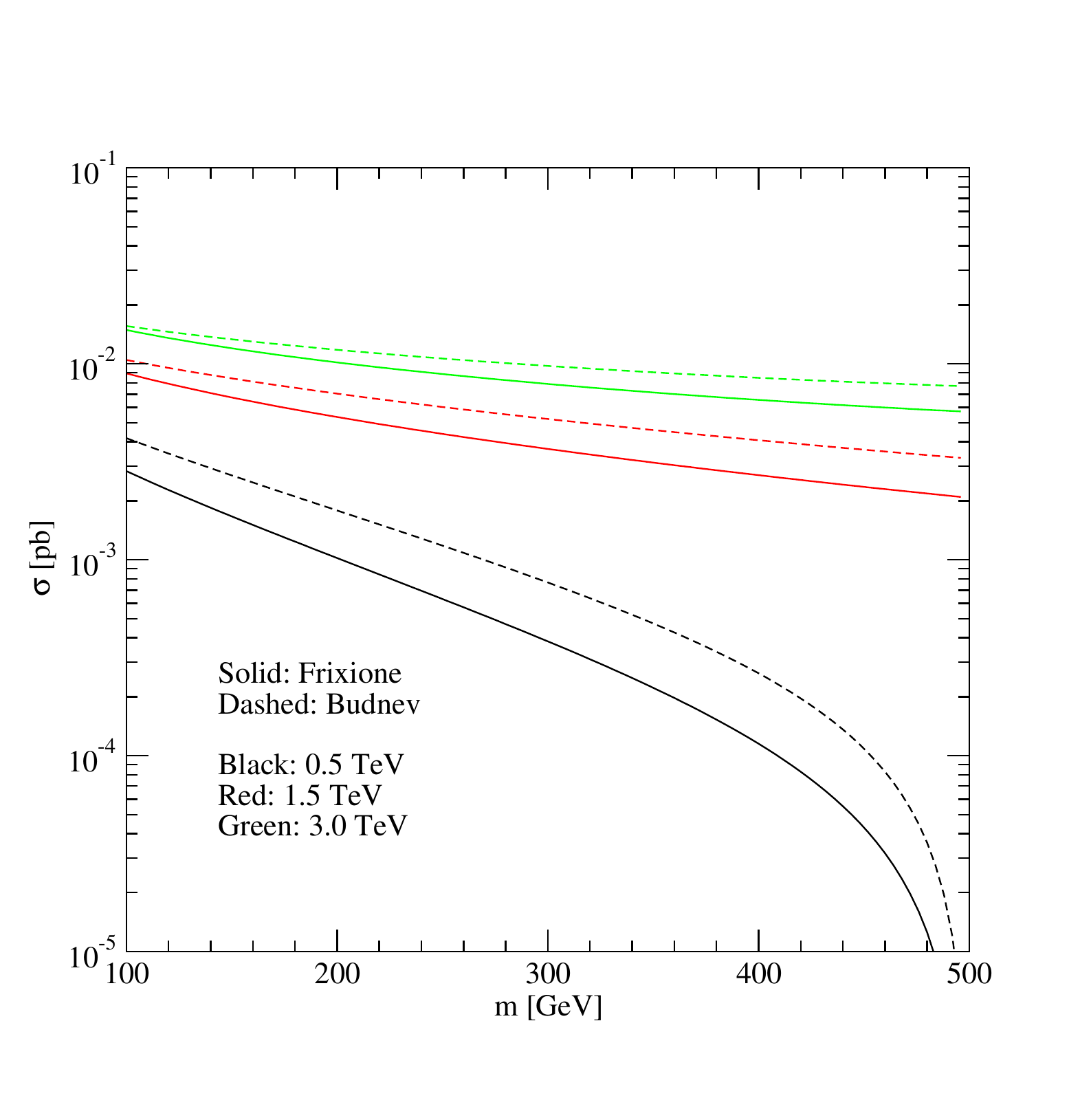}
\end{center}
 \caption{Cross section for the production of technipions as function of the mass, for different energies and photon luminosities, see legend in figure. The left plot is $M_Q = \unit[300]{GeV}$ and right plot is $M_Q = \unit[500]{GeV}$.} 
\end{figure}

The result for cross section of production of technipions as function of mass is displayed in the Fig. (\ref{fig:tech}).

\section{Conclusions}
In this work, estimates are presented for the cross section of BSM particles production in electron-positron linear colliders, expected to start into operation in the coming years. These colliders offer a unique opportunity to study these exotic states in a clear way.

This process is unfavorable in relation to the energy of center of mass and the photon luminosity of the hadron collisors, where it is possible to emit a large number of photons, available for interaction and later creation of states. But, as mentioned above, in larges values of $x$, the electron luminosity surpass the hadron luminosity. In electron collisions, the final state is cleaner and also, due to the absence of (known) electron structure, we do not have the possibility of multiple final states in the detector, generated by the dissociation of hadronic projectiles.

The results for the cross section presented depend, in general, strongly on the mass of the produced particle ($\sigma \sim m^3$), resulting, as expected, in small cross sections for large masses. In general, we have higher results when Budnev photon luminosity, Eq. (\ref{eq:bud}), is used than Frixione expression, Eq. (\ref{eq:frix}). It is also noted that the half-life of the produced states can be very short, and it is not possible to measure the state produced in the detector so only the products of decay can be measured. However, as we are considering different models, there are other free parameters that make the cross section measurable.

\begin{acknowledgments}
The author thanks the Grupo de Altas e Médias Energias (IFM-UFPel) for the support in all stages of this work.
\end{acknowledgments}

\bibliography{referencias}

\begin{thebibliography}{32}
\expandafter\ifx\csname natexlab\endcsname\relax\def\natexlab#1{#1}\fi
\expandafter\ifx\csname bibnamefont\endcsname\relax
  \def\bibnamefont#1{#1}\fi
\expandafter\ifx\csname bibfnamefont\endcsname\relax
  \def\bibfnamefont#1{#1}\fi
\expandafter\ifx\csname citenamefont\endcsname\relax
  \def\citenamefont#1{#1}\fi
\expandafter\ifx\csname url\endcsname\relax
  \def\url#1{\texttt{#1}}\fi
\expandafter\ifx\csname urlprefix\endcsname\relax\def\urlprefix{URL }\fi
\providecommand{\bibinfo}[2]{#2}
\providecommand{\eprint}[2][]{\url{#2}}

\bibitem[{\citenamefont{de~Blas et~al.}(2018)\citenamefont{de~Blas,
  Franceschini, Riva, Roloff, Schnoor, Spannowsky, Wells, Wulzer, Zupan,
  Alipour-Fard et~al.}}]{1812.02093v2}
\bibinfo{author}{\bibfnamefont{J.}~\bibnamefont{de~Blas}},
  \bibinfo{author}{\bibfnamefont{R.}~\bibnamefont{Franceschini}},
  \bibinfo{author}{\bibfnamefont{F.}~\bibnamefont{Riva}},
  \bibinfo{author}{\bibfnamefont{P.}~\bibnamefont{Roloff}},
  \bibinfo{author}{\bibfnamefont{U.}~\bibnamefont{Schnoor}},
  \bibinfo{author}{\bibfnamefont{M.}~\bibnamefont{Spannowsky}},
  \bibinfo{author}{\bibfnamefont{J.~D.} \bibnamefont{Wells}},
  \bibinfo{author}{\bibfnamefont{A.}~\bibnamefont{Wulzer}},
  \bibinfo{author}{\bibfnamefont{J.}~\bibnamefont{Zupan}},
  \bibinfo{author}{\bibfnamefont{S.}~\bibnamefont{Alipour-Fard}},
  \bibnamefont{et~al.} (\bibinfo{year}{2018}), \eprint{1812.02093v2},
  \urlprefix\url{http://arxiv.org/abs/1812.02093v2;
  http://arxiv.org/pdf/1812.02093v2}.

\bibitem[{\citenamefont{Gori}(2015)}]{1511.00405v1}
\bibinfo{author}{\bibfnamefont{S.}~\bibnamefont{Gori}},
  \emph{\bibinfo{title}{{Exploration of Beyond the Standard Model Physics at
  the ILC}}} (\bibinfo{year}{2015}), \eprint{1511.00405v1},
  \urlprefix\url{http://arxiv.org/abs/1511.00405v1;
  http://arxiv.org/pdf/1511.00405v1}.

\bibitem[{\citenamefont{Reis and Sauter}(2017)}]{Reis:2017rvb}
\bibinfo{author}{\bibfnamefont{J.~T.} \bibnamefont{Reis}} \bibnamefont{and}
  \bibinfo{author}{\bibfnamefont{W.~K.} \bibnamefont{Sauter}},
  \bibinfo{journal}{Phys. Rev.} \textbf{\bibinfo{volume}{D96}},
  \bibinfo{pages}{075031} (\bibinfo{year}{2017}), \eprint{1707.04170}.

\bibitem[{\citenamefont{Goncalves and
  Sauter}(2010)}]{goncalves2010RadproexcproCERLHC}
\bibinfo{author}{\bibfnamefont{V.}~\bibnamefont{Goncalves}} \bibnamefont{and}
  \bibinfo{author}{\bibfnamefont{W.}~\bibnamefont{Sauter}},
  \bibinfo{journal}{Phys.Rev.} \textbf{\bibinfo{volume}{D82}},
  \bibinfo{pages}{056009} (\bibinfo{year}{2010}), \eprint{1007.5487}.

\bibitem[{\citenamefont{Goncalves and Sauter}(2015)}]{Goncalves:2015oua}
\bibinfo{author}{\bibfnamefont{V.~P.} \bibnamefont{Goncalves}}
  \bibnamefont{and} \bibinfo{author}{\bibfnamefont{W.~K.}
  \bibnamefont{Sauter}}, \bibinfo{journal}{Phys. Rev.}
  \textbf{\bibinfo{volume}{D91}}, \bibinfo{pages}{035004}
  (\bibinfo{year}{2015}), \eprint{1501.06354}.

\bibitem[{\citenamefont{Nystrand}(2005)}]{nystrand2005Eleintnucprocol}
\bibinfo{author}{\bibfnamefont{J.}~\bibnamefont{Nystrand}},
  \bibinfo{journal}{Nucl.Phys.} \textbf{\bibinfo{volume}{A752}},
  \bibinfo{pages}{470} (\bibinfo{year}{2005}), \eprint{hep-ph/0412096}.

\bibitem[{\citenamefont{Frixione et~al.}(1993)\citenamefont{Frixione, Mangano,
  Nason, and Ridolfi}}]{frixione1993ImpWeiappele-procol}
\bibinfo{author}{\bibfnamefont{S.}~\bibnamefont{Frixione}},
  \bibinfo{author}{\bibfnamefont{M.~L.} \bibnamefont{Mangano}},
  \bibinfo{author}{\bibfnamefont{P.}~\bibnamefont{Nason}}, \bibnamefont{and}
  \bibinfo{author}{\bibfnamefont{G.}~\bibnamefont{Ridolfi}},
  \bibinfo{journal}{Phys.Lett.} \textbf{\bibinfo{volume}{B319}},
  \bibinfo{pages}{339} (\bibinfo{year}{1993}), \eprint{hep-ph/9310350}.

\bibitem[{\citenamefont{Budnev et~al.}(1975)\citenamefont{Budnev, Ginzburg,
  Meledin, and Serbo}}]{budnev1975TwophoparpromecPhyproAppEquphoapp}
\bibinfo{author}{\bibfnamefont{V.}~\bibnamefont{Budnev}},
  \bibinfo{author}{\bibfnamefont{I.}~\bibnamefont{Ginzburg}},
  \bibinfo{author}{\bibfnamefont{G.}~\bibnamefont{Meledin}}, \bibnamefont{and}
  \bibinfo{author}{\bibfnamefont{V.}~\bibnamefont{Serbo}},
  \bibinfo{journal}{Phys.Rept.} \textbf{\bibinfo{volume}{15}},
  \bibinfo{pages}{181} (\bibinfo{year}{1975}).

\bibitem[{\citenamefont{Drees et~al.}(1994)\citenamefont{Drees, Godbole,
  Nowakowski, and Rindani}}]{drees1994gamgamprohigppcol}
\bibinfo{author}{\bibfnamefont{M.}~\bibnamefont{Drees}},
  \bibinfo{author}{\bibfnamefont{R.~M.} \bibnamefont{Godbole}},
  \bibinfo{author}{\bibfnamefont{M.}~\bibnamefont{Nowakowski}},
  \bibnamefont{and} \bibinfo{author}{\bibfnamefont{S.~D.}
  \bibnamefont{Rindani}}, \bibinfo{journal}{Phys.Rev.}
  \textbf{\bibinfo{volume}{D50}}, \bibinfo{pages}{2335} (\bibinfo{year}{1994}),
  \eprint{hep-ph/9403368}.

\bibitem[{\citenamefont{Greiner and Reinhardt}(2009)}]{greinerqed}
\bibinfo{author}{\bibfnamefont{W.}~\bibnamefont{Greiner}} \bibnamefont{and}
  \bibinfo{author}{\bibfnamefont{J.}~\bibnamefont{Reinhardt}},
  \emph{\bibinfo{title}{{Quantum Electrodynamics}}}
  (\bibinfo{publisher}{Springer}, \bibinfo{year}{2009}), ISBN
  \bibinfo{isbn}{978-3540875604}.

\bibitem[{\citenamefont{Klusek-Gawenda and
  Szczurek}(2010)}]{KlusekGawenda:2010kx}
\bibinfo{author}{\bibfnamefont{M.}~\bibnamefont{Klusek-Gawenda}}
  \bibnamefont{and} \bibinfo{author}{\bibfnamefont{A.}~\bibnamefont{Szczurek}},
  \bibinfo{journal}{Phys. Rev.} \textbf{\bibinfo{volume}{C82}},
  \bibinfo{pages}{014904} (\bibinfo{year}{2010}), \eprint{1004.5521}.

\bibitem[{\citenamefont{Godunov et~al.}(2020)\citenamefont{Godunov, Novikov,
  Rozanov, Vysotsky, and Zhemchugov}}]{godunov2020quasistable}
\bibinfo{author}{\bibfnamefont{S.}~\bibnamefont{Godunov}},
  \bibinfo{author}{\bibfnamefont{V.}~\bibnamefont{Novikov}},
  \bibinfo{author}{\bibfnamefont{A.}~\bibnamefont{Rozanov}},
  \bibinfo{author}{\bibfnamefont{M.}~\bibnamefont{Vysotsky}}, \bibnamefont{and}
  \bibinfo{author}{\bibfnamefont{E.}~\bibnamefont{Zhemchugov}},
  \bibinfo{journal}{Journal of High Energy Physics}
  \textbf{\bibinfo{volume}{2020}}, \bibinfo{pages}{143} (\bibinfo{year}{2020}).

\bibitem[{\citenamefont{Randall and
  Sundrum}(1999)}]{randall1999Larmashiesmaextdim}
\bibinfo{author}{\bibfnamefont{L.}~\bibnamefont{Randall}} \bibnamefont{and}
  \bibinfo{author}{\bibfnamefont{R.}~\bibnamefont{Sundrum}},
  \bibinfo{journal}{Phys.Rev.Lett.} \textbf{\bibinfo{volume}{83}},
  \bibinfo{pages}{3370} (\bibinfo{year}{1999}), \eprint{hep-ph/9905221}.

\bibitem[{\citenamefont{Bae et~al.}(2000)\citenamefont{Bae, Ko, Lee, and
  Lee}}]{bae2000PheradRanscecol}
\bibinfo{author}{\bibfnamefont{S.}~\bibnamefont{Bae}},
  \bibinfo{author}{\bibfnamefont{P.}~\bibnamefont{Ko}},
  \bibinfo{author}{\bibfnamefont{H.~S.} \bibnamefont{Lee}}, \bibnamefont{and}
  \bibinfo{author}{\bibfnamefont{J.}~\bibnamefont{Lee}},
  \bibinfo{journal}{Phys.Lett.} \textbf{\bibinfo{volume}{B487}},
  \bibinfo{pages}{299} (\bibinfo{year}{2000}), \eprint{hep-ph/0002224}.

\bibitem[{\citenamefont{Cheung}(2001)}]{cheung2001PheradRansce}
\bibinfo{author}{\bibfnamefont{K.-m.} \bibnamefont{Cheung}},
  \bibinfo{journal}{Phys.Rev.} \textbf{\bibinfo{volume}{D63}},
  \bibinfo{pages}{056007} (\bibinfo{year}{2001}), \eprint{hep-ph/0009232}.

\bibitem[{\citenamefont{Morrissey et~al.}(2012)\citenamefont{Morrissey, Plehn,
  and Tait}}]{ScienceDirect:S037015731200083X}
\bibinfo{author}{\bibfnamefont{D.~E.} \bibnamefont{Morrissey}},
  \bibinfo{author}{\bibfnamefont{T.}~\bibnamefont{Plehn}}, \bibnamefont{and}
  \bibinfo{author}{\bibfnamefont{T.~M.~P.} \bibnamefont{Tait}},
  \bibinfo{journal}{Physics Reports} \textbf{\bibinfo{volume}{515}},
  \bibinfo{pages}{1} (\bibinfo{year}{2012}),
  \urlprefix\url{https://www.sciencedirect.com/science/article/pii/S037015731200083X}.

\bibitem[{\citenamefont{Clark et~al.}(1987)\citenamefont{Clark, Leung, and
  Love}}]{clark1987ProDil}
\bibinfo{author}{\bibfnamefont{T.}~\bibnamefont{Clark}},
  \bibinfo{author}{\bibfnamefont{C.~N.} \bibnamefont{Leung}}, \bibnamefont{and}
  \bibinfo{author}{\bibfnamefont{S.}~\bibnamefont{Love}},
  \bibinfo{journal}{Phys.Rev.} \textbf{\bibinfo{volume}{D35}},
  \bibinfo{pages}{997} (\bibinfo{year}{1987}).

\bibitem[{\citenamefont{Barger et~al.}(2012{\natexlab{a}})\citenamefont{Barger,
  Ishida, and Keung}}]{barger2012DilLHC}
\bibinfo{author}{\bibfnamefont{V.}~\bibnamefont{Barger}},
  \bibinfo{author}{\bibfnamefont{M.}~\bibnamefont{Ishida}}, \bibnamefont{and}
  \bibinfo{author}{\bibfnamefont{W.-Y.} \bibnamefont{Keung}},
  \bibinfo{journal}{Phys.Rev.} \textbf{\bibinfo{volume}{D85}},
  \bibinfo{pages}{015024} (\bibinfo{year}{2012}{\natexlab{a}}),
  \eprint{1111.2580}.

\bibitem[{\citenamefont{Barger et~al.}(2012{\natexlab{b}})\citenamefont{Barger,
  Ishida, and Keung}}]{barger2012DifHigbosdilradhadcol}
\bibinfo{author}{\bibfnamefont{V.}~\bibnamefont{Barger}},
  \bibinfo{author}{\bibfnamefont{M.}~\bibnamefont{Ishida}}, \bibnamefont{and}
  \bibinfo{author}{\bibfnamefont{W.-Y.} \bibnamefont{Keung}},
  \bibinfo{journal}{Phys.Rev.Lett.} \textbf{\bibinfo{volume}{108}},
  \bibinfo{pages}{101802} (\bibinfo{year}{2012}{\natexlab{b}}),
  \eprint{1111.4473}.

\bibitem[{\citenamefont{Coleppa et~al.}(2012)\citenamefont{Coleppa, Gregoire,
  and Logan}}]{coleppa2012DilconLHCpro}
\bibinfo{author}{\bibfnamefont{B.}~\bibnamefont{Coleppa}},
  \bibinfo{author}{\bibfnamefont{T.}~\bibnamefont{Gregoire}}, \bibnamefont{and}
  \bibinfo{author}{\bibfnamefont{H.~E.} \bibnamefont{Logan}},
  \bibinfo{journal}{Phys.Rev.} \textbf{\bibinfo{volume}{D85}},
  \bibinfo{pages}{055001} (\bibinfo{year}{2012}), \eprint{1111.3276}.

\bibitem[{\citenamefont{Coriano et~al.}(2013)\citenamefont{Coriano, {Delle
  Rose}, Quintavalle, and Serino}}]{coriano2013DilintanobrescainvStaMod}
\bibinfo{author}{\bibfnamefont{C.}~\bibnamefont{Coriano}},
  \bibinfo{author}{\bibfnamefont{L.}~\bibnamefont{{Delle Rose}}},
  \bibinfo{author}{\bibfnamefont{A.}~\bibnamefont{Quintavalle}},
  \bibnamefont{and} \bibinfo{author}{\bibfnamefont{M.}~\bibnamefont{Serino}},
  \bibinfo{journal}{JHEP} \textbf{\bibinfo{volume}{1306}}, \bibinfo{pages}{077}
  (\bibinfo{year}{2013}), \eprint{1206.0590}.

\bibitem[{\citenamefont{Chacko et~al.}(2013)\citenamefont{Chacko, Franceschini,
  and Mishra}}]{chacko2013Res125GeVHigorDil}
\bibinfo{author}{\bibfnamefont{Z.}~\bibnamefont{Chacko}},
  \bibinfo{author}{\bibfnamefont{R.}~\bibnamefont{Franceschini}},
  \bibnamefont{and} \bibinfo{author}{\bibfnamefont{R.~K.}
  \bibnamefont{Mishra}}, \bibinfo{journal}{JHEP}
  \textbf{\bibinfo{volume}{1304}}, \bibinfo{pages}{015} (\bibinfo{year}{2013}),
  \eprint{1209.3259}.

\bibitem[{\citenamefont{Bellazzini et~al.}(2013)\citenamefont{Bellazzini,
  Csaki, Hubisz, Serra, and Terning}}]{bellazzini2013HigDil}
\bibinfo{author}{\bibfnamefont{B.}~\bibnamefont{Bellazzini}},
  \bibinfo{author}{\bibfnamefont{C.}~\bibnamefont{Csaki}},
  \bibinfo{author}{\bibfnamefont{J.}~\bibnamefont{Hubisz}},
  \bibinfo{author}{\bibfnamefont{J.}~\bibnamefont{Serra}}, \bibnamefont{and}
  \bibinfo{author}{\bibfnamefont{J.}~\bibnamefont{Terning}},
  \bibinfo{journal}{Eur.Phys.J.} \textbf{\bibinfo{volume}{C73}},
  \bibinfo{pages}{2333} (\bibinfo{year}{2013}), \eprint{1209.3299}.

\bibitem[{\citenamefont{Jung and Ko}(2014)}]{jung2014HigsysconLHCHigdat}
\bibinfo{author}{\bibfnamefont{D.-W.} \bibnamefont{Jung}} \bibnamefont{and}
  \bibinfo{author}{\bibfnamefont{P.}~\bibnamefont{Ko}},
  \bibinfo{journal}{Phys.Lett.} \textbf{\bibinfo{volume}{B732}},
  \bibinfo{pages}{364} (\bibinfo{year}{2014}), \eprint{1401.5586}.

\bibitem[{\citenamefont{Ebadi et~al.}(2019)\citenamefont{Ebadi, Khatibi, and
  {Mohammadi Najafabadi}}}]{Ebadi:2019gij}
\bibinfo{author}{\bibfnamefont{J.}~\bibnamefont{Ebadi}},
  \bibinfo{author}{\bibfnamefont{S.}~\bibnamefont{Khatibi}}, \bibnamefont{and}
  \bibinfo{author}{\bibfnamefont{M.}~\bibnamefont{{Mohammadi Najafabadi}}},
  \bibinfo{journal}{Phys. Rev.} \textbf{\bibinfo{volume}{D100}},
  \bibinfo{pages}{015016} (\bibinfo{year}{2019}), \eprint{1901.03061}.

\bibitem[{\citenamefont{Bauer et~al.}(2019)\citenamefont{Bauer, Heiles,
  Neubert, and Thamm}}]{Bauer:2018uxu}
\bibinfo{author}{\bibfnamefont{M.}~\bibnamefont{Bauer}},
  \bibinfo{author}{\bibfnamefont{M.}~\bibnamefont{Heiles}},
  \bibinfo{author}{\bibfnamefont{M.}~\bibnamefont{Neubert}}, \bibnamefont{and}
  \bibinfo{author}{\bibfnamefont{A.}~\bibnamefont{Thamm}},
  \bibinfo{journal}{Eur. Phys. J.} \textbf{\bibinfo{volume}{C79}},
  \bibinfo{pages}{74} (\bibinfo{year}{2019}), \eprint{1808.10323}.

\bibitem[{\citenamefont{Bauer et~al.}(2017)\citenamefont{Bauer, Neubert, and
  Thamm}}]{Bauer:2017ris}
\bibinfo{author}{\bibfnamefont{M.}~\bibnamefont{Bauer}},
  \bibinfo{author}{\bibfnamefont{M.}~\bibnamefont{Neubert}}, \bibnamefont{and}
  \bibinfo{author}{\bibfnamefont{A.}~\bibnamefont{Thamm}},
  \bibinfo{journal}{JHEP} \textbf{\bibinfo{volume}{12}}, \bibinfo{pages}{044}
  (\bibinfo{year}{2017}), \eprint{1708.00443}.

\bibitem[{\citenamefont{Rybka}(2017)}]{Rybka:2017swh}
\bibinfo{author}{\bibfnamefont{G.}~\bibnamefont{Rybka}}, \bibinfo{journal}{J.
  Phys.} \textbf{\bibinfo{volume}{G44}}, \bibinfo{pages}{124002}
  (\bibinfo{year}{2017}).

\bibitem[{\citenamefont{Inan and Kisselev}(2020)}]{2003.01978v1}
\bibinfo{author}{\bibfnamefont{S.~C.} \bibnamefont{Inan}} \bibnamefont{and}
  \bibinfo{author}{\bibfnamefont{A.~V.} \bibnamefont{Kisselev}},
  \emph{\bibinfo{title}{{A search for axion-like particles in light-by-light
  scattering at the CLIC}}} (\bibinfo{year}{2020}), \eprint{2003.01978v1},
  \urlprefix\url{http://arxiv.org/abs/2003.01978v1;
  http://arxiv.org/pdf/2003.01978v1}.

\bibitem[{\citenamefont{Knapen et~al.}(2017)\citenamefont{Knapen, Lin, Lou, and
  Melia}}]{Knapen:2016moh}
\bibinfo{author}{\bibfnamefont{S.}~\bibnamefont{Knapen}},
  \bibinfo{author}{\bibfnamefont{T.}~\bibnamefont{Lin}},
  \bibinfo{author}{\bibfnamefont{H.~K.} \bibnamefont{Lou}}, \bibnamefont{and}
  \bibinfo{author}{\bibfnamefont{T.}~\bibnamefont{Melia}},
  \bibinfo{journal}{Phys. Rev. Lett.} \textbf{\bibinfo{volume}{118}},
  \bibinfo{pages}{171801} (\bibinfo{year}{2017}), \eprint{1607.06083}.

\bibitem[{\citenamefont{Baldenegro et~al.}(2018)\citenamefont{Baldenegro,
  Fichet, von Gersdorff, and Royon}}]{Baldenegro:2018hng}
\bibinfo{author}{\bibfnamefont{C.}~\bibnamefont{Baldenegro}},
  \bibinfo{author}{\bibfnamefont{S.}~\bibnamefont{Fichet}},
  \bibinfo{author}{\bibfnamefont{G.}~\bibnamefont{von Gersdorff}},
  \bibnamefont{and} \bibinfo{author}{\bibfnamefont{C.}~\bibnamefont{Royon}},
  \bibinfo{journal}{JHEP} \textbf{\bibinfo{volume}{06}}, \bibinfo{pages}{131}
  (\bibinfo{year}{2018}), \eprint{1803.10835}.

\bibitem[{\citenamefont{Lebiedowicz et~al.}(2014)\citenamefont{Lebiedowicz,
  Pasechnik, and Szczurek}}]{lebiedowicz2014SeatecexcprodiplarinvmasLHC}
\bibinfo{author}{\bibfnamefont{P.}~\bibnamefont{Lebiedowicz}},
  \bibinfo{author}{\bibfnamefont{R.}~\bibnamefont{Pasechnik}},
  \bibnamefont{and} \bibinfo{author}{\bibfnamefont{A.}~\bibnamefont{Szczurek}},
  \bibinfo{journal}{Nucl.Phys.} \textbf{\bibinfo{volume}{B881}},
  \bibinfo{pages}{288} (\bibinfo{year}{2014}), \eprint{1309.7300}.

\end{thebibliography}

\end{document}